\documentstyle[12pt,epsf]{article}
\newcounter{sectionc}\newcounter{subsectionc}\newcounter{subsubsectionc}
\renewcommand{\section}[1] {\vspace{0.6cm}\addtocounter{sectionc}{1}
\setcounter{subsectionc}{0}\setcounter{subsubsectionc}{0}\noindent
	{\bf\thesectionc. #1}\par\vspace{0.4cm}}
\renewcommand{\subsection}[1] {\vspace{0.6cm}\addtocounter{subsectionc}{1}
	\setcounter{subsubsectionc}{0}\noindent
	{\it\thesectionc.\thesubsectionc. #1}\par\vspace{0.4cm}}
\renewcommand{\subsubsection}[1]
{\vspace{0.6cm}\addtocounter{subsubsectionc}{1}
	\noindent {\rm\thesectionc.\thesubsectionc.\thesubsubsectionc.
	#1}\par\vspace{0.4cm}}
\newcommand{\nonumsection}[1] {\vspace{0.6cm}\noindent{\bf #1}
	\par\vspace{0.4cm}}

\renewenvironment{thebibliography}[1]
	{\begin{list}{\arabic{enumi}.}
	{\usecounter{enumi}\setlength{\parsep}{0pt}
\setlength{\leftmargin 1.25cm}{\rightmargin 0pt}
	 \setlength{\itemsep}{0pt} \settowidth
	{\labelwidth}{#1.}\sloppy}}{\end{list}}

 \catcode`@=11
\def\@cite#1#2{\unskip\nobreak\relax
    \def\@tempa{$\m@th^{\hbox{\the\scriptfont0 #1}}$}%
    \futurelet\@tempc\@citexx}
\def\@citexx{\ifx.\@tempc\let\@tempd=\@citepunct\else
    \ifx,\@tempc\let\@tempd=\@citepunct\else
    \let\@tempd=\@tempa\fi\fi\@tempd}
\def\@citepunct{\@tempc\edef\@sf{\spacefactor=\the\spacefactor\relax}\@tempa
    \@sf\@gobble}

\def\citenum#1{{\def\@cite##1##2{##1}\cite{#1}}}
\def\citea#1{\@cite{#1}{}}

\newcount\@tempcntc
\def\@citex[#1]#2{\if@filesw\immediate\write\@auxout{\string\citation{#2}}\fi
  \@tempcnta\z@\@tempcntb\m@ne\def\@citea{}\@cite{\@for\@citeb:=#2\do
    {\@ifundefined
       {b@\@citeb}{\@citeo\@tempcntb\m@ne\@citea\def\@citea{,}{\bf ?}\@warning
       {Citation `\@citeb' on page \thepage \space undefined}}%
    {\setbox\z@\hbox{\global\@tempcntc0\csname b@\@citeb\endcsname\relax}%
     \ifnum\@tempcntc=\z@ \@citeo\@tempcntb\m@ne
       \@citea\def\@citea{,}\hbox{\csname b@\@citeb\endcsname}%
     \else
      \advance\@tempcntb\@ne
      \ifnum\@tempcntb=\@tempcntc
      \else\advance\@tempcntb\m@ne\@citeo
      \@tempcnta\@tempcntc\@tempcntb\@tempcntc\fi\fi}}\@citeo}{#1}}
\def\@citeo{\ifnum\@tempcnta>\@tempcntb\else\@citea\def\@citea{,}%
  \ifnum\@tempcnta=\@tempcntb\the\@tempcnta\else
   {\advance\@tempcnta\@ne\ifnum\@tempcnta=\@tempcntb \else \def\@citea{--}\fi
    \advance\@tempcnta\m@ne\the\@tempcnta\@citea\the\@tempcntb}\fi\fi}

\newcommand{\tcaption}[1]{
        \refstepcounter{table}
        \setbox\@tempboxa = \hbox{\tenrm Table~\thetable. #1}
        \ifdim \wd\@tempboxa > 6in
           {\begin{center}
        \parbox{6in}{\tenrm\baselineskip=12pt Table~\thetable. #1}
            \end{center}}
        \else
             {\begin{center}
             {\tenrm Table~\thetable. #1}
              \end{center}}
        \fi}
\catcode`@=12

\def\lsim{\mathrel{\raise.2ex\hbox{$<$}\hskip-.8em\lower.9ex\hbox{$\sim$}}}
\def\gsim{\mathrel{\raise.2ex\hbox{$>$}\hskip-.8em\lower.9ex\hbox{$\sim$}}}

\begin{document}
\thispagestyle{empty}
\baselineskip14pt
\parindent=1.5pc

\font\fortssbx=cmssbx10 scaled \magstep1
\hbox to \hsize{
\includegraphics{uwlogo.ps}
\hskip.25in \raise.1in\hbox{\fortssbx University of Wisconsin - Madison}
\hfill$\vbox{\hbox{\bf MADPH-95-902}
                \hbox{July 1995}}$ }

\begin{center}
\hbox{\small\bf
The Direct and Indirect Detection of Weakly Interacting Dark Matter
Particles\footnotemark}\smallskip
\small
F. Halzen\\
\footnotesize\it
University of Wisconsin, Department of Physics, Madison WI 53706\\
and\\
University of Hawaii, Department of Physics, Honolulu HI 96822
\end{center}
\footnotetext{Talk presented at the {\it International Symposium on Particle
Theory and Phenomenology},\break Iowa State University, May 22--24, 1995.}

\vspace*{-.17in}

\begin{center}
\footnotesize
ABSTRACT
\smallskip

\parbox{6in}{\baselineskip13pt
An ever-increasing body of evidence suggests that weakly interacting massive
particles (WIMPs) constitute the bulk of the matter in the Universe. We
illustrate how experimental data, dimensional analysis and Standard Model
particle physics are sufficient to evaluate and compare the potential of
detectors searching for such particles either directly (e.g.\ by their
scattering in germanium detectors), or indirectly (e.g.\ by observing their
annihilation into neutrinos in underground detectors).}
\end{center}

\vskip-.15in

\section{Introduction and Results}

It has become widely accepted that most of our Universe is made of cold dark
matter particles. Big bang cosmology implies that these particles have
interactions of order the weak scale, i.e.\ they are WIMPs\cite{Seckel}. In the
early Universe WIMPs are in equilibrium with photons. When the Universe cools
to temperatures well below the mass $m_\chi$ of the WIMP their density is
Boltzmann-suppressed as $\exp(-m_\chi/T)$ and would, today, be exponentially
small if it were not for the expansion of the Universe. At some point, as a
result of this expansion, WIMPs drop out of equilibrium with other particles
and a relic abundance persists. The mechanism is analogous to nucleosynthesis
where the density of helium and other elements is determined by competition
between the rate of nuclear reactions and the expansion of the Universe.

For WIMPs to make up a large fraction of the Universe today, i.e.\ a large
fraction of $\Omega$, their annihilation cross section has to be ``just
right''. The annihilation cross section can be dimensionally written as $
\alpha^2/m_\chi^2$, where $\alpha$ is the fine-structure constant. It then
follows that
\begin{equation}
	\Omega\propto1/\sigma\propto m_\chi^2 \,. \label{omega}
\end{equation}
The critical point is that for $\Omega\simeq1$ we find that $m_\chi\simeq m_W$,
the mass of the weak intermediate boson. There is a deep connection between
critical cosmological density and the weak scale. Weakly interacting particles
which constitute the bulk of the mass of the Universe remain to be discovered.
When our galaxy was formed the cold dark matter inevitably clustered with the
luminous matter to form a sizeable fraction of the
\begin{equation}
        \rho_{\chi}=0.4\rm~GeV/cm^3  \label{density}
\end{equation}
galactic matter density implied by observed rotation curves. Unlike the
baryons, the dissipationless WIMPs fill the galactic halo which is believed to
be an isothermal{\parfillskip0pt\par\noindent} sphere of WIMPs with average
velocity
\begin{equation}
         v_{\chi}=300\rm\ km/sec \,. \label{velocity}
\end{equation}

In summary, we know everything about these particles (except whether they
really exist!). We know that their mass is of order of the weak boson mass
with:
\begin{equation}
       \mbox{tens of GeV} < m_{\chi} < \rm several\ TeV \,. \label{GT}
\end{equation}
Lower masses are excluded by accelerator and (in)direct searches while masses
beyond several TeV are excluded by cosmological considerations. We know that
WIMPs interact weakly. We also know their density and average velocity in our
Galaxy given the assumption that they constitute the dominant component of the
density of our galactic halo as measured by rotation curves.

Two general techniques, referred to as direct (D) and indirect (ID),  are
pursued to demonstrate the existence of WIMPs\cite{Seckel}. In direct detectors
one observes the energy deposited when WIMPs elastically scatter off nuclei.
The indirect method infers the existence of WIMPs from observation of their
annihilation products. WIMPs will annihilate into neutrinos; massive WIMPs will
annihilate into high-energy neutrinos which can be detected in high-energy
neutrino telescopes. Throughout this paper we will assume that such neutrinos
are detected in a generic Cherenkov detector which measures the direction and,
to some extent, the energy of a secondary muon produced by a neutrino of WIMP
origin in or near the instrument\cite{Gaisser}. It can also detect the showers
initiated by electron-neutrinos.

The indirect detection is greatly facilitated by the fact that the sun
represents a dense and nearby source of accumulated cold dark matter
particles\cite{Drees}. Galactic WIMPs, scattering off nuclei in the sun, lose
energy. They may fall below escape velocity and be gravitationally trapped.
Trapped WIMPs eventually come to equilibrium temperature and accumulate near
the center of the sun. While the WIMP density builds up, their annihilation
rate into lighter particles increases until equilibrium is achieved where the
annihilation rate equals half of the capture rate. The sun has thus become a
reservoir of WIMPs which we expect to annihilate mostly into heavy quarks and,
for the heavier WIMPs, into weak bosons. The leptonic decays of the heavy quark
and weak boson annihilation products turn the sun into a source of high-energy
neutrinos with energies in the GeV to TeV range.

The performance of future detectors is determined by the rate of elastic
scattering of WIMPs in a low-background, germanium detector and, for the
indirect method, by the flux of solar neutrinos of WIMP origin. Both are a
function of WIMP mass and of their elastic cross section on nucleons. In
standard cosmology WIMP capture and annihilation interactions are weak, and we
will suggest that, given this constraint, dimensional analysis is sufficient to
compute the scattering rates in germanium detectors as well as the neutrino
flux from the measured WIMP density in our galactic halo. We will derive and
compare rates for direct and indirect detection of weakly interacting particles
with mass $m_\chi \simeq m_W$ assuming

\begin{enumerate}
\advance\itemsep by -0.05in
\item
that WIMPs represent the major fraction of the measured halo density. Their
flux is
\begin{equation}
\phi_\chi = n_\chi v_\chi = {0.4\over m_\chi} \, {\rm {GeV\over cm^3} \
3\times10^6 {cm\over s} } = {1.2\times10^7\over m_{\chi\rm\,GeV}} \,\rm
cm^{-2} s^{-1} \;,
\label{phi chi}
\end{equation}
where $m_{\chi\rm\,GeV} \equiv (m_\chi/$1~GeV) is in GeV units.

\item
a WIMP-nucleon interaction cross section based on dimensional analysis
\begin{equation}
\sigma(\chi N) = \left(G_F m_N^2\right)^2 {1\over m_W^2} \equiv \sigma_{\rm DA}
= 6\times10^{-42}\rm\,cm^2 \;,
\label{sigma chi N}
\end{equation}

\item
that WIMPs annihilate 10\% of the time in neutrinos (this is just the leptonic
branching ratio of the final state particles in the dominant annihilation
channels $\chi\bar\chi \to W^+W^-$ or $Q\bar Q$, where $Q$ is a heavy quark).

\end{enumerate}

Clearly the cross section for the interaction of WIMPs with matter is
uncertain. Arguments can be invoked to raise or decrease it. Important points
are that i) our choice represents a typical intermediate value, ii) all our
results for event rates scale linearly in the cross section and can be easily
reinterpreted, and iii) the comparison of direct and indirect event rates is
independent of the choice.

We present a simple and totally transparent analysis in which the event rates
of detectors are derived from the above assumptions. It finesses all detailed
dynamics and gives answers that are sufficiently accurate considering that the
mass of the particle has not been pinned down. We will find that the event rate
in a direct detector is proportional to the WIMP cross section and flux and the
density of targets $m_N^{-1}$, i.e.
\begin{equation}
{dN_{\rm D} \over dM} = {1\over m_N} \phi_\chi \sigma_{\rm DA} N(A_D) =
{1.4\over m_{\chi\rm\,GeV}} \rm\ (kg)^{-1} \, (year)^{-1} \nonumber\\,
\label{direct}
\end{equation}
where ${dN_{\rm D}\over dM}$ represents the number of direct events per unit of
target mass. $N(A_D$) represents the coherent enhancement factor for a nuclear
target of atomic number $A_D$, e.g.\ 76 for Germanium,
\begin{equation}
N(A) \equiv A^3 \left[ 1 + {m_\chi\over m_N} \over A + {m_\chi\over m_N}
\right]^2 \;.
\label{nuclear}
\end{equation}
The rates for indirect detection are
\begin{equation}
dN_{\rm ID}/ dA \simeq \left\{ 1.8\times10^{-2}m_{\chi\rm\,GeV} \right\}
\left\{ \rho^{\vphantom0}(A_{ID}) N(A_{ID})\right\} \left\{ 1+1.9\times10^{-4}
m_{\chi\rm\,GeV} \right\}^{-7} \;,
\label{indirect}
\end{equation}
where $dN_{\rm ID} / dA$, in units of $\rm\ (10^4\,m^2)^{-1} (year)^{-1}$,
represents the number of events from the sun per unit area $A$ detected by a
neutrino telescope. The factor \{$\rho N$\} should be summed over all elements
in the sun. Because of additional nuclear form factor effects which are
neglected in Eq.~\ref{indirect} it is adequate to consider oxygen with a solar
abundance of $\rho =1.1$~\% and $A_{ID} = 16$ as a ``typical'' element. The
observed average muon energy should be in the range $1/4 \sim 1/6
m_{\chi\rm\,GeV}$.

The above parametrizations readily lead to the conclusion that the direct
method is superior if the WIMP interacts coherently on nuclei (which has been
assumed for Eqs.~\ref{direct}--\ref{indirect}) and, if its mass is lower or
comparable to the weak boson mass $m_W$. We will show that in all other cases,
i.e.\ for relatively heavy WIMPs and for all WIMPs interacting incoherently,
the indirect method is competitive or superior, but it is, of course, held
hostage to the successful deployment of high energy neutrino telescopes with
effective area in the $\sim10^4$--$10^6$~m$^2$ range and with appropriately low
threshold. Especially for heavier WIMPs the indirect technique is powerful
because underground high energy neutrino detectors have been optimized to be
sensitive in the energy region where the neutrino interaction cross section and
the range of the muon are large. A kilometer-size detector probes WIMP masses
up to the TeV-range, beyond which they are excluded by cosmological
considerations.

For high energy neutrinos the muon and neutrino are aligned, with good angular
resolution, along a direction pointing back to the sun. The number of
background events of atmospheric neutrino origin in the pixel containing the
signal will be small. The angular spread of secondary muons from neutrinos
coming from the direction of the sun is well described by the
relation\cite{Gaisser} $\sim 1.2^\circ \Big/ \sqrt{E_\mu(\rm TeV)}$.
Measurement of muon energy, which may be only up to order of magnitude accuracy
in some experiments, can be used to infer the WIMP mass from the angular spread
of the signal. The spread contains information on the neutrino energy and,
therefore, the WIMP mass. More realistically, measurement of the muon energy
can be used to reduce the search window around the sun, resulting in a reduced
background.

\looseness=-1
Before proceeding, we comment on our ansatz for the elastic WIMP-nucleon
scattering cross section. The simplest dimensional analysis implies that the
cross section is $G_F^2 m_N^2$. This correctly describes the $Z$-exchange
diagram of Fig~1a, which is of the form
\begin{equation}
\sigma\ \sim\ G_F^2 {m_N^2m_\chi^2\over (m_N+m_\chi)^2} \;.
\label{sigd}
\end{equation}

\vskip-.1in

\begin{figure}[h]
\centering

\epsfxsize=3.25in\hspace{0in}\epsffile{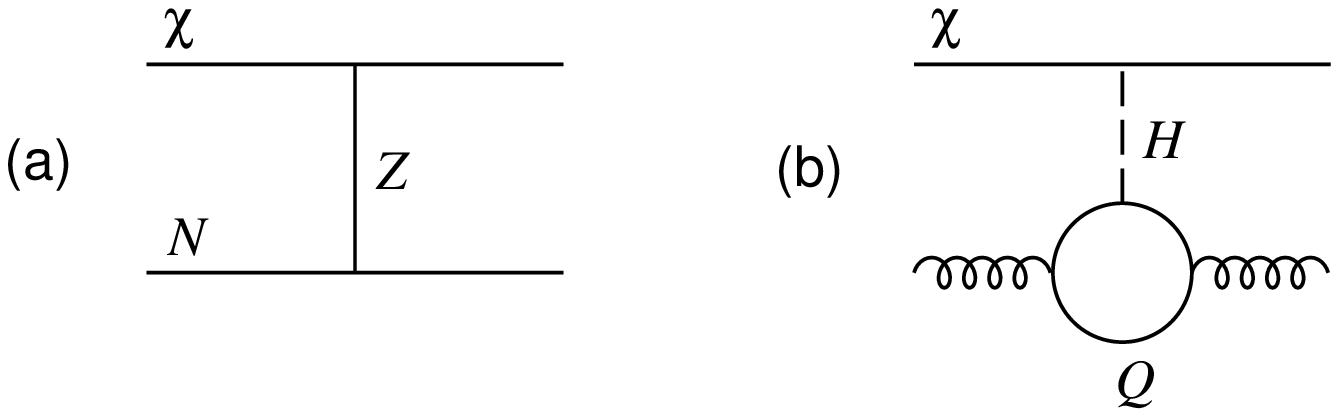}

\medskip
\parbox{6in}{\footnotesize\baselineskip13pt Fig.~1. Examples of (a) incoherent
and (b)~coherent WIMP-nucleon interactions. In (b) the gluon is a constituent
of the target nucleon and $Q$ is a heavy quark.}

\end{figure}

For coherent interactions, which we will emphasize throughout this paper, there
is an additional suppression factor associated with the exchange of the Higgs
particle with a mass of order of the weak boson mass; see Fig~1b. In the
specific diagram shown the Higgs interacts with the heavy quarks in the gluon
condensate associated with the nucleon target. It is of the form
\begin{equation}
\sigma \sim G_F g_H^2 {m_N^2 m_\chi^2\over(m_N+m_\chi)^2} {1\over m_W^2} \;,
\label{sigid}
\end{equation}
where $g_H \sim \sqrt{G_F}\, m_N$  describes the condensate. Conservatively, we
will use the suppressed WIMP interaction cross section which is appropriate for
coherent scattering.

\section{Derivation of Detection rates}

For the case of direct detection the structure of Eq.~\ref{direct} is
transparent\cite{Seckel}. For indirect detection the number of solar neutrinos
of WIMP origin can be calculated in 5 easy steps by determining:

\begin{itemize}
\advance\itemsep by -0.05in

\item
the capture cross section in the sun, which is given by the product of the
number of target nucleons in the sun and the elastic scattering cross section
\begin{equation}
\sigma_\odot = f  \left[ 1.2\times10^{57} \right] \sigma_{\rm DA} \;.
\label{sigma sun}
\end{equation}
This includes a focussing factor $f$ given, as usual, by the ratio of kinetic
and potential energy of the WIMP near the sun. It enhances the capture rate by
a factor 10.

\item
the WIMP flux from the sun which is given by
\begin{equation}
\phi_\odot = \phi_\chi \sigma_\odot / 4\pi d^2 \;,
\label{phi sun}
\end{equation}
where $d=1\rm~a.u. = 1.5\times10^{13}\,cm$.

\item
the actual neutrino flux, which is obtained after inclusion of the branching
ratio. From (\ref{phi chi}),(\ref{sigma chi N}) and (\ref{sigma sun}),(\ref{phi
sun})
\begin{equation}
\phi_\nu = 10^{-1} \times \phi_\odot
= {3\times10^{-5}\over m_{\chi\rm\,GeV}}\rm \, cm^{-2} \, s^{-1} \;.
\end{equation}

\item
the probability to detect the neutrino\cite{Gaisser}, which is proportional to
\begin{eqnarray}
&&P = \rho\sigma_\nu R_\mu,\rm\ with\nonumber\\
&&\rho = \mbox{Avagadro\,\#} = 6\times10^{23}\nonumber\\
&&\sigma_\nu = \mbox{neutrino interaction cross section} = 0.5\times10^{-38}\
E_\nu\rm (GeV)\ cm^{2}\nonumber\\
&&R_\mu = \mbox{muon range} = 500{\rm\ cm}\ E_\mu\rm(GeV)\, \nonumber\\
\noalign{\vskip2pt}
\rm or&&P = 2\times 10^{-13} \, m_{\chi\rm\,GeV}^2
\end{eqnarray}
Here we assumed the kinematics of the decay chain
\begin{eqnarray}
\chi\bar\chi &\to& W^+W^- \nonumber \\  \noalign{\vskip-1ex}
             &   & \hspace{2em} \raise1ex\hbox{$\vert$}\!{\rightarrow}
\mu\nu_\mu \nonumber
\end{eqnarray}
with $E_\nu = {1\over2}m_\chi$ (this would be ${1\over3}m_\chi$ for $Q$ decay)
and $E_\mu = {1\over2}E_\nu = {m_\chi\over 4}$.

\item
finally, $dN_{\rm ID} / dA = \phi_{\nu} P = 1.8\times10^{-6} \,
m_{\chi\rm\,GeV} \, \rm\ (year)^{-1} \, (m^2)^{-1}$
\stepcounter{equation}\hfill(\theequation)\break
where $dN_{\rm ID} / dA$ represents the number of events from the sun per unit
area (m$^2$) detected by a neutrino telescope.

\end{itemize}

\noindent

We can now summarize our results so far by comparing a $10^4$~m$^2$ neutrino
detector, an area typical of the instruments now being deployed, with a
kilogram of hydrogen:
\begin{eqnarray}
dN_{\rm ID}/ dA &=& 1.8\times10^{-2} m_{\chi\rm\,GeV} \rm\ (10^4\,m^2)^{-1}
(year)^{-1} \nonumber\\
dN_{\rm D}/ dM &=& {1.4\over m_{\chi\rm\,GeV}} \rm\ (kg)^{-1} \, (year)^{-1}
\nonumber\\
{dN_{\rm D}/dM\over dN_{\rm ID}/dA} \left(10^4\rm\,m^2\over\rm kg\right)
&=& {7.8\times10^1\over m_{\chi\rm\,GeV}^2}  \label{D/ID}
\end{eqnarray}
Direct detection is superior only in the mass range $m_\chi<10$~GeV, but this
region is, arguably, ruled out by previous searches. Indirect detection is the
preferred technique. This straightforward conclusion may, however, be
invalidated when WIMPs interact coherently with nuclei and targets other than
hydrogen are considered. We discuss this next.

\section{Coherent Nuclear Enhancements}

For WIMPs interacting coherently with nuclei in the detector or in the sun, the
nuclear dependence of the event rates resides in

\begin{itemize}
\advance\itemsep by -0.1in
\item
the target density factor $m_N^{-1}$ in Eq.~\ref{direct} which is replaced by
$(Am_N)^{-1}$,

\item
the coherent enhancement factor ``$A^2$",

\item
the nuclear dependence of the cross sections of Eqs.~\ref{sigd},\ref{sigid}
which is obtained by the substitution
\begin{eqnarray}
\noalign{\qquad\underline{\rm incoherent}}
\sigma &\sim& G_F^2 {m_N^2m_\chi^2\over (m_N+m_\chi)^2} \to G_F^2  {(Am_N)^2
m_\chi^2\over (Am_N+m_\chi)^2} \nonumber\\ 
\noalign{\qquad\underline{\rm coherent}}
\sigma &\sim& G_F^2 g_H^2 {m_N^2 m_\chi^2\over(m_N+m_\chi)^2} {1\over m_W^2}
\to G_F\left(g_H\over m_W\right)^2 {(Am_N)^2 m_\chi^2\over (Am_N+m_\chi)^2} A^2
\nonumber
\end{eqnarray}
The coherent enhancement factor for a nucleus $A$, including a factor $A^{-1}$
for the target density,  is therefore given by
\begin{equation}
{1\over A} {A^2 (Am_N)^2 m_\chi^2 \over (Am_N+m_\chi)^2 } \,
{(m_N+m_\chi)^2\over m_N^2 m_\chi^2}
= A^3 {(m_N+m_\chi)^2\over (Am_N+m_\chi)^2}
= A^3\left[ 1+{m_\chi\over m_N} \over A+{m_\chi\over m_N} \right]^2 .
\end{equation}

\end{itemize}
This yields Eq.~\ref{nuclear}.

\section{Event Rates for WIMPs with Coherent Interactions}

Our simple evaluations, made so far, overestimate the indirect rates for very
heavy WIMPS because high energy neutrinos, created by annihilation near the
core, may be absorbed in the sun. Absorption is stronger for neutrinos and,
therefore, mostly antineutrinos form the signature for very heavy WIMPS. The
probability that an antineutrino escapes without absorption is well
parametrized by $(1+ 3.8 \times 10^{-4} E_{\nu})^{-7}$, where $E_{\nu} \simeq
m_{\chi}/2$. This yields our final result of Eq.~\ref{indirect}.

The relative merits of the two methods are illustrated in the following table,
which is obtained from Eqs.~\ref{direct}--\ref{indirect} and establishes that a
kilogram of germanium and a $10^4$~m$^2$ are competitive.

\vskip-.3in

\begin{table}[h]
\tcaption{Event rates and signal to background $(N/B)$.}
\centering\unskip\smallskip
\tabcolsep=1.5em
\begin{tabular}{c|c@{\quad}cc@{\quad}c}
\hline
$m_\chi$ (GeV)&\multicolumn{2}{c}{Direct (/kg/year)}&
\multicolumn{2}{c}{Indirect (/$10^4$\,m$^2$/year)}\\
\hline
& events& $N/B$& events& $N/B$\\
50 & $2.2\times10^3$& 7& $2.3\times10^1$& $\simeq\,1$\\
500 & $1.1\times10^3$& 7& $2\times10^2$& $\simeq\,10^2$\\
2000 & $2.9\times10^2$& 1& $1.7\times10^2$& $\simeq\,10^4$\\
\hline
\end{tabular}
\end{table}

At the lower energy the event rates for the indirect method are underestimated
because also the Earth is a source of neutrinos of WIMP origin.

We conclude that the direct method yields more events for the lower masses,
even when compared to a $10^6$~m$^2$ detector. As expected, the indirect method
is competitive for heavier WIMPs with a detection rate growing like $E_\nu^2$
or $m_\chi^2$. A $10^5$~m$^2$ instrument covers the full WIMP mass range, even
if the WIMPs do not coherently interact with nuclei in the sun. These
conclusions are reinforced after considering the signal-to-noise for both
techniques which we discuss next.

\section{Backgrounds}

\noindent
\underline{Indirect Background}.
For the indirect detection the background event rate is determined by the flux
of atmospheric neutrinos in the detector coming from a pixel around the
sun\cite{Gaisser}. The number of events in a $10^4$~m$^2$ detector is $\sim
10^2/E_\mu$(TeV) and the pixel size is determined by the angle between muon and
neutrino $\sim 1.2^\circ \Big/ \sqrt{E_\mu(\rm TeV)}$. Using the kinematics
$E_\mu \simeq m_\chi/4$ we obtain
\[
B_{\rm ID} = { 10^2/E_\mu({\rm TeV}) \over 2\pi \Big/
\left[ 1.2^\circ {\pi\over 180^\circ} \over \sqrt{E_\mu(\rm TeV)} \right]^2}
= {1.1\times10^5\over m_{\chi\rm\,GeV}^2} \mbox{ per 10$^4$\,m$^2$ per year}
\]
This is only valid for large $m_\chi$, i.e.\ for $E_\mu \cong m_\chi/4
>100$~GeV. Without this approximation we obtain

\vskip-.11in

\begin{table}[h]
\centering\unskip
\tabcolsep=1.25em
\begin{tabular}{c|ccc}
\hline
 &\small  \# bkgd. events&\small \# pixels of solar&\small bkgd. events\\
\noalign{\vskip-.85ex}
 &\small in 10$^4$\,m$^2$&\small size in $2\pi$&\small per 10$^4$\,m$^2$\\
\noalign{\vskip-.85ex}
$E_\mu$(GeV) &\small in $2\pi$ &\small  &\small per pixel, per year \\
\hline
10& 3200& 140& 23\\
100& 1060& $1.4\times10^3$& 0.8\\
1000& 110& $1.4\times10^4$& $8\times10^{-3}$\\
\hline
\end{tabular}
\end{table}

\noindent
For large $m_\chi$ the signal to background ratio is
\[
\left(N\over B\right)_{\rm ID} \equiv {dN_{\rm ID}/dA\over dB_{\rm ID}/dA}
\simeq
7.2\times10^{-6} m_{\chi\rm\,GeV}^3
\]
Clearly, the extremely optimistic predictions for signal-to-noise are unlikely
to survive the realities of experimental physics. One expects, typically, to
measure muon energy only to order-of-magnitude accuracy in the initial
experiments. The energy of showers initiated by electron neutrinos should be
determined to a factor 2. It is not excluded that future, dedicated experiments
may do better. The conclusion that high energy muons pointing at the sun
represents a superb signature, is unlikely to be invalidated.

\smallskip
\noindent
\underline{Direct Background}: about 300 events per year per
kg\cite{Kamionkowski}. Signal-to-noise therefore exceeds unity up to 2~TeV WIMP
mass.

These considerations were used to estimate the signal-to-noise $N/B$ in
Table~1.

\section{Dynamics?}

We emphasize that our results are representative for the specific and much
studied example where the lightest supersymmetric particle is Nature's
WIMP\cite{Drees}. Clearly dynamics, which is now defined, can alter our
conclusions, but only in ``conspiratorial" ways. Dynamics can, on the other
hand, increase rates as well, sometimes by well over an order of magnitude,
over and above the rates obtained from dimensional analysis in this paper. Our
qualitative conclusions are valid, at least in some average sense, in
supersymmetry. Our results do, in fact, closely trace the supersymmetry
prediction of reference 2 for the choice of Higgs coupling ${g_H^2 \over 4
\pi}=1$, in their notation.

We feel that the development of detectors should be guided by an analysis like
ours rather than by dynamics of theories beyond the standard model for which
there is, at present, no experimental guidance.

\bigskip

\leftline{\bf Acknowledgements}
\medskip

We thank M.~Drees, S.~Pakvasa and X.~Tata for a careful reading of the
manuscript.
This research was supported in part by the U.S.~Department of Energy under
Grant No.~DE-FG02-95ER40896 and in part by the University of Wisconsin Research
Committee with funds granted by the Wisconsin Alumni Research Foundation.

\nonumsection{References}\unskip

\end{document}